\documentstyle[amsfonts,12pt]{article}
\begin{document}

\title{Quantum control and the Strocchi map}
\author{{\it R.Vilela Mendes}\thanks{%
corresponding author: vilela@cii.fc.ul.pt} \\
%EndAName
{\small Complexo Interdisciplinar, Universidade de Lisboa}\\
{\small Av. Prof. Gama Pinto, 2, 1699 Lisboa Codex, Portugal} \and {\it V.
I. Man'ko} \\
%EndAName
{\small P. N. Lebedev Physical Institute }\\
{\small Leninsky Prospect 53, 117924 Moscow, Russia}}
\date{}
\maketitle

\begin{abstract}
Identifying the real and imaginary parts of wave functions with coordinates
and momenta, quantum evolution may be mapped onto a classical Hamiltonian
system. In addition to the symplectic form, quantum mechanics also has a
positive-definite real inner product which provides a geometrical
interpretation of the measurement process. Together they endow the quantum
Hilbert space with the structure of a K\"{a}ller manifold.

Quantum control is discussed in this setting. Quantum time-evolution
corresponds to smooth Hamiltonian dynamics and measurements to jumps in the
phase space. This adds additional power to quantum control, non unitarily
controllable systems becoming controllable by ``measurement plus evolution''.

A picture of quantum evolution as Hamiltonian dynamics in a classical-like
phase-space is the appropriate setting to carry over techniques from
classical to quantum control. This is illustrated by a discussion of optimal
control and sliding mode techniques.
\end{abstract}

\section{Introduction}

The mathematical structures of classical and quantum mechanics are usually
regarded as essentially different. However, many years ago Strocchi\cite
{Strocchi}, by identifying the real and imaginary parts of the wave function
with coordinates and momenta, has shown that quantum evolution may be mapped
onto a classical Hamiltonian system. In particular this setting suits nicely
a geometrical interpretation of quantum mechanics. This formulation of
quantum mechanics, first proposed by Strocchi\cite{Strocchi}, has since been
rediscovered and extended by many authors\cite{Kibble} \cite{Heslot} \cite
{Anadan} \cite{Cirelli} \cite{Ashtekar}. Some applications of the Strocchi
map to the evolution of finite-dimensional quantum systems were considered
in \cite{Marmo2} and structures relevant to the relation of classical to
quantum mechanics were studied in \cite{Marmo1}. A particularly interesting
extension to the original ideas of the Strocchi map was the realization
that, in addition to the symplectic form, characteristic of Hamiltonian
evolution, quantum mechanics also has a positive-definite real inner
product. Together they endow the quantum Hilbert space with the structure of
a K\"{a}ller manifold\cite{Heslot} \cite{Ashtekar}. We will discuss this
more general framework of the Strocchi map. However, for simplicity and
whenever possible, we will adhere to the original intuitive coordinate
formulation of Strocchi.

The aim of this work is to discuss quantum control in the geometrical
setting provided by the Strocchi map. Quantum time-evolution will correspond
to smooth Hamiltonian dynamics in a classical-like phase-space and
measurements to jumps in the phase-space. This is very different from the
situation in classical feedback control, where the measurements needed for
the feedback action are not supposed to change the state of the system or to
change it only very little. However, in some cases, rather than being a
nuisance, the state disturbance introduced by quantum measurement adds
additional power to quantum feedback control and, in particular, it changes
the question of controllability. Non unitarily controllable systems may
become controllable by ``measurement plus evolution''.

The paper is organized as follows: In section~2 a review is made of the
properties of the Strocchi map as well as some extensions needed to describe
quantum evolution and control in both the von Neumann and
positive-operator-valued measure approaches. In section~3, a new quantum
control method is proposed which combines free quantum evolution and quantum
measurement to reach a desirable quantum state. Finally, in section~4,
quantum analogs of classical nonlinear and optimal control methods are
shortly discussed.

\section{Geometrical formulation of quantum mechanics. The Strocchi map}

Consider a basis $\left\{ \left| k\right\rangle \right\} $ in a separable
complex Hilbert space ${\cal H}^{*}$. A general quantum state $\left| \psi
\right\rangle $ is 
\begin{equation}
\left| \psi \right\rangle =\sum_{k}\psi _{k}\left| k\right\rangle
\label{1.1}
\end{equation}
Define 
\begin{equation}
\psi _{k}=\frac{1}{\sqrt{2}}\left( q_{k}+ip_{k}\right)  \label{1.2}
\end{equation}
where $\left\{ q_{k},p_{k}\right\} $ is a numerable set of real phase-space
coordinates. Then the scalar product in the complex Hilbert space ${\cal H}%
^{*}$ 
\begin{equation}
\left\langle \psi ^{^{\prime }}|\psi \right\rangle =\frac{1}{2}%
\sum_{k}\left( q_{k}^{^{\prime }}q_{k}+p_{k}^{^{\prime }}p_{k}\right)
+i\left( q_{k}^{^{\prime }}p_{k}-p_{k}^{^{\prime }}q_{k}\right) =\frac{1}{2}%
\left\{ G\left( \psi ^{^{\prime }},\psi \right) +i\Omega \left( \psi
^{^{\prime }},\psi \right) \right\}  \label{1.3}
\end{equation}
decomposes into the sum of a positive real inner product 
\begin{equation}
G\left( \psi ^{^{\prime }},\psi \right) =\frac{1}{2}\sum_{k}\left(
q_{k}^{^{\prime }}q_{k}+p_{k}^{^{\prime }}p_{k}\right)  \label{1.3a}
\end{equation}
and a symplectic form 
\begin{equation}
\Omega \left( \psi ^{^{\prime }},\psi \right) =\frac{1}{2}\sum_{k}\left(
q_{k}^{^{\prime }}p_{k}-p_{k}^{^{\prime }}q_{k}\right)  \label{1.3b}
\end{equation}

Considering ${\cal H}^{*}=\left( {\cal H},J\right) $ as a real Hilbert space 
${\cal H}$ with a complex structure $J$, the triple $\left( J,G,\Omega
\right) $ equips ${\cal H}$ with the structure of a K\"{a}hler space because 
\begin{equation}
G\left( \psi ^{^{\prime }},\psi \right) =\Omega \left( \psi ^{^{\prime
}},J\psi \right)  \label{1.4}
\end{equation}

The Schr\"{o}dinger equation $i\frac{\partial }{\partial t}\left| \psi
\right\rangle =H\left| \psi \right\rangle $ becomes the set of Hamilton's
equations 
\begin{equation}
\begin{array}{lll}
\frac{d}{dt}q_{k} & = & \frac{\partial }{\partial p_{k}}{\Bbb H} \\ 
\frac{d}{dt}p_{k} & = & -\frac{\partial }{\partial q_{k}}{\Bbb H}
\end{array}
\label{1.5}
\end{equation}
associated to the symplectic form $\Omega \left( \psi ^{^{\prime }},\psi
\right) $ and the ``classical'' Hamiltonian 
\begin{equation}
{\Bbb H}=\frac{1}{2}\sum_{k,j}\left\{ \left( q_{k}q_{j}+p_{k}p_{j}\right) 
\textnormal{Re}H_{kj}+\left( p_{k}q_{j}-q_{k}p_{j}\right) \textnormal{Im}%
H_{kj}\right\}  \label{1.6}
\end{equation}
with $H_{kj}=\left\langle k|H|j\right\rangle $.

One sees that the time evolution of quantum mechanics is equivalent to the
classical dynamics of a numerable set of coupled oscillators. What is unique
to quantum mechanics is the special role played by the symmetric form $%
G\left( \psi ^{^{\prime }},\psi \right) $.

Let ${\cal S}$ be the Hilbert sphere, that is, the space of normalized $%
\left( \left\| \psi \right\| =1\right) $ functions in the Hilbert space $%
{\cal H}$. $G\left( \psi ^{^{\prime }},\psi \right) $ defines a metric in $%
{\cal S}$. Consider now a measurement of an observable $A$ which, for
simplicity, we assume to have a (possibly degenerate) discrete spectrum. Let 
$a$ be a (degenerate) eigenvalue of $A$ and $P_{a}$ the projector on the
subspace $V_{a}$ of ${\cal S}$ associated to this eigenvalue. After the
measurement of $A$ is performed and the value is found to be $a$ , the
quantum state changes from $\psi \in {\cal S}$ to $\psi _{a}=\frac{P_{a}\psi 
}{\left\| P_{a}\psi \right\| }\in {\cal S}$ and the probability to find this
value is $\left\| P_{a}\psi \right\| ^{2}$. The metric $G\left( \psi
^{^{\prime }},\psi \right) $ provides a nice geometrical interpretation of
the measurement process in quantum mechanics.

Given $\psi \in {\cal S}$ and $\phi \in V_{a}\subset {\cal S}$ it is easy to
see that 
\begin{equation}
\left( \psi -\phi ,\psi -\phi \right)  \label{1.7}
\end{equation}
is minimal when $\phi =\psi _{a}$. Because $\left( \psi -\phi ,\psi -\phi
\right) =G\left( \psi -\phi ,\psi -\phi \right) $ one concludes that the
measurement projects $\psi $ on the element of $V_{a}$ that is closest to $%
\psi $ in the $G-$metric. The probability for this projection is 
\begin{equation}
p_{a}=\left\| P_{a}\psi \right\| ^{2}=\left( 1-\frac{1}{2}G\left( \psi -%
\frac{P_{a}\psi }{\left\| P_{a}\psi \right\| },\psi -\frac{P_{a}\psi }{%
\left\| P_{a}\psi \right\| }\right) \right) ^{2}  \label{1.8}
\end{equation}
Therefore, whereas the symplectic form $\Omega $ determines time-evolution,
the $G-$metric controls the measurement process. It is the special role
played by the metric that, in this framework, sets apart quantum from
classical mechanics.

In the numerable basis $\left\{ \left| k\right\rangle \right\} $ of finite
or infinite cardinality $\chi $, each pure quantum state is represented by a
point $\left( \overrightarrow{q},\overrightarrow{p}\right) $ in a
``phase-space'' of dimension $2\chi $. Similarly a mixed state will be
described by a density on the same space. The density matrix for a mixed
state of the form $\rho \left( t\right) =\sum_{n}\rho _{n}\left| \psi
_{n}\left( t\right) \right\rangle \left\langle \psi _{n}\left( t\right)
\right| $ $\left( \sum_{n}\rho _{n}=1\right) $ becomes, using the notations
in Eq.(\ref{1.2})

\begin{equation}
\rho \left( t\right) =\int d\overrightarrow{q}d\overrightarrow{p}\rho \left( 
\overrightarrow{q},\overrightarrow{p}\right) \sum_{k,k^{^{\prime }}}\left(
q_{k}\left( t\right) +ip_{k}\left( t\right) \right) \left( q_{k^{^{\prime
}}}\left( t\right) -ip_{k^{^{\prime }}}\left( t\right) \right) \left|
k\right\rangle \left\langle k^{^{\prime }}\right|  \label{1.9}
\end{equation}
Using the equations of motion (\ref{1.5}) and integration by parts one
obtains the following classical-like equation for the ``phase-space''
density 
\begin{equation}
\frac{d}{dt}\rho \left( \overrightarrow{q},\overrightarrow{p}\right) =-\frac{%
\partial \rho }{\partial \overrightarrow{q}}\cdot \frac{\partial {\Bbb H}}{%
\partial \overrightarrow{p}}+\frac{\partial \rho }{\partial \overrightarrow{p%
}}\cdot \frac{\partial {\Bbb H}}{\partial \overrightarrow{q}}=-\left\{ \rho ,%
{\Bbb H}\right\}  \label{1.10}
\end{equation}

In the Strocchi map framework, the (unobserved) dynamics of quantum states
is a continuous symplectic evolution in a phase space. On the other hand,
measurement of a state is represented by jumps in phase space. Because the
basis $\left\{ \left| k\right\rangle \right\} $ is arbitrary, we may suppose
that this is a basis of eigenstates of the set ${\cal K}$ of observables
that is being measured. Let the state before the measurement be $\left( 
\overrightarrow{q},\overrightarrow{p}\right) $. When a measurement is
performed and the results registered to be $k$, the state jumps from $\left( 
\overrightarrow{q},\overrightarrow{p}\right) $ to $\left( \overrightarrow{q}=%
\frac{q_{k}}{\sqrt{q^{2}+p^{2}}}\overrightarrow{e_{k}},\overrightarrow{p}=%
\frac{p_{k}}{\sqrt{q^{2}+p^{2}}}\overrightarrow{e_{k}^{^{\prime }}}\right) $%
where $\overrightarrow{e_{k}}$ and $\overrightarrow{e_{k}^{^{\prime }}}$ are
unit vectors along the $k-$coordinate and the $k-$momentum.

In feedback control, the results of a measurement are used to correct the
driving forces and different corrections will be associated to different
results of the measurement. Therefore the kind of measurement of interest in
quantum control is a selective one, that is, one in which the result of the
measurement is registered. However, in quantum mechanics, even if the
results of the measurement are no registered, the state of the system (or
our information about it) is changed anyway. For such non-selective
measurements one obtains a mixed state. If, for example the initial state is
a normalized pure state, corresponding to the phase space vector $\left( 
\overrightarrow{q},\overrightarrow{p}\right) $, after the measurement the
state corresponds to a phase-space density 
\begin{equation}
\rho \left( \overrightarrow{\mu },\overrightarrow{\nu }\right)
=\sum_{k}\left( q_{k}^{2}+p_{k}^{2}\right) \delta \left( \overrightarrow{\mu 
}-\frac{q_{k}}{\sqrt{q^{2}+p^{2}}}\overrightarrow{e_{k}}\right) \delta
\left( \overrightarrow{\nu }-\frac{p_{k}}{\sqrt{q^{2}+p^{2}}}\overrightarrow{%
e_{k}^{^{\prime }}}\right)  \label{1.11}
\end{equation}

The above considerations refer to complete quantum mechanical projections,
that is to (selective or non-selective) outputs of quantum mechanical
measurements. A description of the behavior of a quantum system under
continuous observation also exists. It uses generalized quantum measurements
implemented as positive operator-valued measures. Caves and Milburn\cite
{Caves} assume that any measurement takes a certain amount of time and that,
in the infinitesimal time interval $dt$, the measurement of the observable $%
\Lambda $ corresponds to the operation 
\begin{equation}
\rho \rightarrow P_{\Lambda }\rho P_{\Lambda }  \label{1.12}
\end{equation}
with 
\begin{equation}
P_{\Lambda }\left( \alpha \right) =\left( \frac{\pi }{2sdt}\right) ^{-\frac{1%
}{4}}e^{-sdt\left( \Lambda -\alpha \right) ^{2}}  \label{1.13}
\end{equation}
and $\int d\alpha P_{\Lambda }^{\dagger }\left( \alpha \right) P_{\Lambda
}\left( \alpha \right) =1$. Applying $P_{\Lambda }\left( \alpha \right) $ to
any state one obtains a superposition of eigenstates of $\Lambda $ with
eigenvectors centered around $\alpha $ and $s$ defines the resolution (or
the strength) of the measurement. Notice however, that $P_{\Lambda }\left(
\alpha \right) $ is not a projection. It is supposed\cite{Caves} to
represent a generalized selective measurement. For a non-selective
measurement, that is, one where the results are not recorded, 
\begin{equation}
\rho \rightarrow \int d\alpha P_{\Lambda }\left( \alpha \right) \rho
P_{\Lambda }\left( \alpha \right)  \label{1.14}
\end{equation}

From Eqs.(\ref{1.13})-(\ref{1.14}) and the unconditional evolution $\frac{%
d\rho }{dt}=-i\left[ H,\rho \right] $ , the following non-selective
continuous observation equation for the density matrix is obtained 
\begin{equation}
\frac{d\rho }{dt}=-i\left[ H,\rho \right] -\frac{s}{2}\left[ \Lambda ,\left[
\Lambda ,\rho \right] \right]  \label{1.15}
\end{equation}
This equation is physically appealing, in the sense that, for example, the
last double commutator term displays a mechanism for the damping of
non-diagonal terms in the density matrix. Notice however that the $%
P_{\Lambda }\left( \alpha \right) $ operations are quantum mechanical
projections only in the limit of infinite strength, or infinite time.

Because the choice of basis is arbitrary we may, without loss of generality,
choose a basis of eigenstates of the measured observable $\Lambda $. Then,
one has the following equation for the matrix elements of the density matrix 
\begin{equation}
\frac{d}{dt}\rho _{kk^{^{\prime }}}=-i\left\langle k\right| \left[ H,\rho
\right] \left| k^{^{\prime }}\right\rangle -\frac{s}{2}\left( \lambda
_{k}-\lambda _{k^{^{\prime }}}\right) ^{2}\rho _{kk^{^{\prime }}}
\label{1.16}
\end{equation}
leading to the damping of non-diagonal terms, $\lambda _{k}$ and $\lambda
_{k^{^{\prime }}}$ being the eigenvalues of $\Lambda $ in the states $\left|
k\right\rangle $ and $\left| k^{^{\prime }}\right\rangle $.

From the Strocchi map ``phase-space'' density point of view, Eq.(\ref{1.16})
means that, if the initial state is a pure state $\rho _{0}\left( 
\overrightarrow{\mu },\overrightarrow{\nu }\right) =\delta \left( 
\overrightarrow{\mu }-\overrightarrow{q}\right) \delta \left( 
\overrightarrow{\nu }-\overrightarrow{p}\right) $, (in addition to the
Hamiltonian evolution) continuous observation splits the density into
several components which, when $t\rightarrow \infty $, converge to a density
as in (\ref{1.11}). Independently of the conceptual interest of generalized
measurements and continuous observation, the important point to retain is
that the operation of measurement modifies Hamiltonian evolution. Hence, it
might play for quantum control a role similar to the one that is played by
dissipation in non-linear classical control techniques.

\section{Control by measurement plus evolution}

As seen before, in the Strocchi map phase-space, undisturbed time-evolution
is a smooth Hamiltonian dynamics in phase-space, whereas measurements
correspond to jumps in the phase-space. This last aspect is very different
from the situation in classical feedback control, where the measurements
needed for the feedback action are not supposed to change the state of the
system or to change it only very little. However in some cases, rather than
being a nuisance, the state disturbance introduced by quantum measurement
adds additional power to quantum feedback control. In particular it changes
the question of controllability.

For quantum systems with Hamiltonian 
\begin{equation}
H\left( t\right) =H_{0}+\sum_{j=1}^{r}u_{j}\left( t\right) H_{j}  \label{3.1}
\end{equation}
Huang, Tarn and Clark\cite{Clark} obtained a general result on
controllability involving the Lie algebras generated by the free and control
Hamiltonians. A bounded quantum system with finite energy has a finite
number $N$ of allowed states. For this case a necessary and sufficient
condition for controllability is that the Lie algebra generated by $\left\{
H_{0,}H_{1},\cdots ,H_{r}\right\} $ be $u\left( N\right) $ \cite{Rama} (or $%
su\left( N\right) $ if phases are not important\cite{Fu}) because this is
the smallest group that acts transitively on the complex sphere $S_{{\Bbb C}%
}^{N-1}$.

When the Strocchi map is used to describe quantum evolution of the $N-$level
system, the phase-space has dimension $2N$ with coordinates $\left\{
q_{k},p_{k}\right\} $. The set of transformations that need to be available
to have controllability is the $U\left( N\right) $ subgroup of $SO\left(
2N\right) $ corresponding to the real linear canonical transformations, that
also preserve the form $G\left( \psi ^{^{\prime }},\psi \right) $ (Eq.(\ref
{1.3a})). It contains $N\left( N-1\right) /2$ one-parameter subgroups of the
type 
\begin{equation}
\begin{array}{lll}
q_{i}^{^{\prime }} & = & q_{i}\cos \theta +q_{j}\sin \theta \\ 
q_{j}^{^{\prime }} & = & -q_{i}\sin \theta +q_{j}\cos \theta \\ 
p_{i}^{^{\prime }} & = & p_{i}\cos \theta +p_{j}\sin \theta \\ 
p_{j}^{^{\prime }} & = & -p_{i}\sin \theta +p_{j}\cos \theta
\end{array}
\label{3.2}
\end{equation}
$N\left( N-1\right) /2$ of type 
\begin{equation}
\begin{array}{lll}
q_{i}^{^{\prime }} & = & q_{i}\cos \theta -p_{j}\sin \theta \\ 
q_{j}^{^{\prime }} & = & -p_{i}\sin \theta +q_{j}\cos \theta \\ 
p_{i}^{^{\prime }} & = & p_{i}\cos \theta +q_{j}\sin \theta \\ 
p_{j}^{^{\prime }} & = & q_{i}\sin \theta +p_{j}\cos \theta
\end{array}
\label{3.3}
\end{equation}
and $N$ of type 
\begin{equation}
\begin{array}{lll}
q_{i}^{^{\prime }} & = & q_{i}\cos \theta -p_{i}\sin \theta \\ 
p_{i}^{^{\prime }} & = & q_{i}\sin \theta +p_{i}\cos \theta
\end{array}
\label{3.4}
\end{equation}

Suppose now that ${\cal A}=\left\{ H_{0,}H_{1},\cdots ,H_{r}\right\} _{LA}$
is a proper subalgebra of $u\left( N\right) $. Then each orbit of the
subgroup $G({\cal A})\subset U\left( N\right) $ does not cover $S_{{\Bbb C}%
}^{N-1}$. $S_{{\Bbb C}}^{N-1}$ becomes a fiber space with the orbits of $G(%
{\cal A})$ as fibers and base $U\left( N\right) /G({\cal A})$. Then a goal
state $\psi _{f}$ can only be reached from $\psi _{0}$ if $\psi _{0}$ and $%
\psi _{f}$ belong to the same fiber. The system is not controllable purely
by the action of the unitary evolution $\int \exp \left( i\tau H(\tau
)\right) d\tau $ but may be controllable by the joint action of {\it %
measurement plus evolution} in the following sense:

{\it Theorem:} Given any goal state $\psi _{f}$ , there is a family of
observables $M\left( \psi _{f}\right) $ such that measurement of one of
these observables on any $\psi _{0}$ plus unitary evolution leads to $\psi
_{f}$ if $G({\cal A})$ is either $O\left( N\right) $ or $Sp\left( \frac{1}{2}%
N\right) $.

{\it Proof:} If $G({\cal A})=O\left( N\right) $ or $Sp\left( \frac{1}{2}%
N\right) $ we may choose an orthonormal basis $\left\{ \phi _{i}\right\} $
for $S^{N-1}$ in the orbit $G({\cal A})\psi _{f}$. Construct an observable $%
M=\sum_{i}a_{i}P_{\phi _{i}}$, $P_{\phi _{i}}$ being the projector on $\phi
_{i}$. Measuring this observable on any state $\psi _{0}$ and recording the
measured value $a_{k}$ the state becomes $\phi _{k}$ and then, by unitary
evolution, $\psi _{f}$ may be reached.

{\it Remarks:}

(i) Because of both the arbitrary nature of the eigenvalues $a_{i}$ and of
the orthonormal basis, there is a large family of observables appropriate
for this type of control.

(ii) In the result above the state $\psi _{f}$ is fixed, but $\psi _{0}$ is
arbitrary. If both $\psi _{0}$ and $\psi _{f}$ are fixed a much simpler set
of controlling interactions $H_{j}$ may be sufficient. $\psi _{0}$ being
fixed, one constructs the $M$ observable by $N-1$ vectors in the $N-1$%
-dimensional subspace orthogonal to $\psi _{0}$ plus a single vector in the
orbit $G({\cal A})\psi _{f}$ , non-orthogonal to $\psi _{0}$. Then, $G({\cal %
A})$ may be a much smaller subgroup than the ones in the theorem.

One sees that, if properly used, the state disturbing effects of quantum
measurement, rather than being a nuisance, add controlling power over
quantum processes.

{\it Examples:}

(i) A simple example of non-controllable 3-level system has been discussed
by Solomon and Schirmer\cite{Solomon1}. Let 
\begin{equation}
H=H_{0}+u\left( t\right) H_{1}  \label{3.4a}
\end{equation}
with 
\begin{equation}
H_{0}=\mu \left( 
\begin{array}{rrr}
-1 & 0 & 0 \\ 
0 & 0 & 0 \\ 
0 & 0 & 1
\end{array}
\right) ;\qquad H_{1}=d\left( 
\begin{array}{rrr}
0 & 1 & 0 \\ 
1 & 0 & 1 \\ 
0 & 1 & 0
\end{array}
\right)  \label{3.5}
\end{equation}
$H_{0}$ and $H_{1}$ generate the algebra of $SO\left( 3\right) $, therefore
the system is not controllable by unitary evolution. The Strocchi map
evolution equations are 
\begin{equation}
\frac{d}{dt}\left( 
\begin{array}{r}
\overrightarrow{q} \\ 
\overrightarrow{p}
\end{array}
\right) =\left( 
\begin{array}{rr}
0 & A \\ 
-A & 0
\end{array}
\right) \left( 
\begin{array}{r}
\overrightarrow{q} \\ 
\overrightarrow{p}
\end{array}
\right)  \label{3.6}
\end{equation}
$A$ being the matrix $\left( 
\begin{array}{ccc}
-\mu & u\left( t\right) d & 0 \\ 
u\left( t\right) d & 0 & u\left( t\right) d \\ 
0 & u\left( t\right) d & \mu
\end{array}
\right) $. Eq.(\ref{3.6}) splits in block form 
\begin{equation}
\frac{d}{dt}\left( 
\begin{array}{r}
\overrightarrow{q+ip} \\ 
\overrightarrow{q-ip}
\end{array}
\right) =i\left( 
\begin{array}{rr}
-A & 0 \\ 
0 & A
\end{array}
\right) \left( 
\begin{array}{r}
\overrightarrow{q+ip} \\ 
\overrightarrow{q-ip}
\end{array}
\right)  \label{3.7}
\end{equation}
exhibiting the $SO\left( 3\right) $ nature of the control. In terms of the
Hilbert space wave functions, Eq.(\ref{3.7}) means that $\psi ^{*}$ cannot
be reached from $\psi $. Three one-parameters subgroups of the $SO\left(
3\right) $ control group, to be used later on, are

$h_{1}\left( \theta \right) :$ 
\begin{eqnarray}
&&\left( 
\begin{array}{r}
\overrightarrow{q^{^{\prime }}} \\ 
\overrightarrow{p^{^{\prime }}}
\end{array}
\right)   \label{3.8a} \\
&=&\left( 
\begin{array}{cccccc}
\frac{1}{2}\left( \cos \theta +1\right)  & 0 & \frac{1}{2}\left( \cos \theta
-1\right)  & 0 & \frac{-1}{\sqrt{2}}\sin \theta  & 0 \\ 
0 & \cos \theta  & 0 & \frac{-1}{\sqrt{2}}\sin \theta  & 0 & \frac{-1}{\sqrt{%
2}}\sin \theta  \\ 
\frac{1}{2}\left( \cos \theta -1\right)  & 0 & \frac{1}{2}\left( \cos \theta
+1\right)  & 0 & \frac{-1}{\sqrt{2}}\sin \theta  & 0 \\ 
0 & \frac{1}{\sqrt{2}}\sin \theta  & 0 & \frac{1}{2}\left( \cos \theta
+1\right)  & 0 & \frac{1}{2}\left( \cos \theta -1\right)  \\ 
\frac{1}{\sqrt{2}}\sin \theta  & 0 & \frac{1}{\sqrt{2}}\sin \theta  & 0 & 
\cos \theta  & 0 \\ 
0 & \frac{1}{\sqrt{2}}\sin \theta  & 0 & \frac{1}{2}\left( \cos \theta
-1\right)  & 0 & \frac{1}{2}\left( \cos \theta +1\right) 
\end{array}
\right) \left( 
\begin{array}{r}
\overrightarrow{q} \\ 
\overrightarrow{p}
\end{array}
\right)   \nonumber
\end{eqnarray}
$h_{2}\left( \theta \right) :$%
\begin{eqnarray}
&&\left( 
\begin{array}{r}
\overrightarrow{q^{^{\prime }}} \\ 
\overrightarrow{p^{^{\prime }}}
\end{array}
\right)   \label{3.9} \\
&=&\left( 
\begin{array}{cccccc}
\frac{1}{2}\left( \cos \theta +1\right)  & \frac{-1}{\sqrt{2}}\sin \theta  & 
\frac{1}{2}\left( 1-\cos \theta \right)  & 0 & 0 & 0 \\ 
\frac{1}{\sqrt{2}}\sin \theta  & \cos \theta  & \frac{-1}{\sqrt{2}}\sin
\theta  & 0 & 0 & 0 \\ 
\frac{1}{2}\left( 1-\cos \theta \right)  & \frac{1}{\sqrt{2}}\sin \theta  & 
\frac{1}{2}\left( \cos \theta +1\right)  & 0 & 0 & 0 \\ 
0 & 0 & 0 & \frac{1}{2}\left( \cos \theta +1\right)  & \frac{-1}{\sqrt{2}}%
\sin \theta  & \frac{1}{2}\left( 1-\cos \theta \right)  \\ 
0 & 0 & 0 & \frac{1}{\sqrt{2}}\sin \theta  & \cos \theta  & \frac{-1}{\sqrt{2%
}}\sin \theta  \\ 
0 & 0 & 0 & \frac{1}{2}\left( 1-\cos \theta \right)  & \frac{1}{\sqrt{2}}%
\sin \theta  & \frac{1}{2}\left( \cos \theta +1\right) 
\end{array}
\right) \left( 
\begin{array}{r}
\overrightarrow{q} \\ 
\overrightarrow{p}
\end{array}
\right)   \nonumber
\end{eqnarray}
$h_{3}\left( \theta \right) :$%
\begin{equation}
\left( 
\begin{array}{r}
\overrightarrow{q^{^{\prime }}} \\ 
\overrightarrow{p^{^{\prime }}}
\end{array}
\right) =\left( 
\begin{array}{cccccc}
\cos \theta  & 0 & 0 & \sin \theta  & 0 & 0 \\ 
0 & 1 & 0 & 0 & 0 & 0 \\ 
0 & 0 & \cos \theta  & 0 & 0 & -\sin \theta  \\ 
-\sin \theta  & 0 & 0 & \cos \theta  & 0 & 0 \\ 
0 & 0 & 0 & 0 & 1 & 0 \\ 
0 & 0 & \sin \theta  & 0 & 0 & \cos \theta 
\end{array}
\right) \left( 
\begin{array}{r}
\overrightarrow{q} \\ 
\overrightarrow{p}
\end{array}
\right)   \label{3.10}
\end{equation}

Although not controllable by unitary evolution alone, the system is
controllable by {\it measurement plus evolution}. Consider, for example,

$\psi _{f}=\left\{ \overrightarrow{q}=\left( 0,0,0\right) ,\overrightarrow{p}%
=\left( \frac{1}{\sqrt{2}},0,\frac{1}{\sqrt{2}}\right) \right\} $ as the
goal state. By applying $h_{1}\left( -\frac{\pi }{2}\right) $ and $%
h_{2}\left( \frac{\pi }{2}\right) h_{1}\left( -\frac{\pi }{2}\right) $ to
this state one obtains an orthogonal set 
\begin{equation}
\begin{array}{l}
\psi _{1}=\left\{ \overrightarrow{q}=\left( 0,1,0\right) ,\overrightarrow{p}%
=\left( 0,0,0\right) \right\} \\ 
\psi _{2}=\left\{ \overrightarrow{q}=\left( \frac{-1}{\sqrt{2}},0,\frac{1}{%
\sqrt{2}}\right) ,\overrightarrow{p}=\left( 0,0,0\right) \right\} \\ 
\psi _{3}=\left\{ \overrightarrow{q}=\left( 0,0,0\right) ,\overrightarrow{p}%
=\left( \frac{1}{\sqrt{2}},0,\frac{1}{\sqrt{2}}\right) \right\}
\end{array}
\label{3.11}
\end{equation}
Denoting by $P_{i}$ the projectors on the states $\psi _{i}$, measurement of
an arbitrary state by any one of the observables in the family 
\begin{equation}
\sum_{i}a_{i}P_{i}  \label{3.12}
\end{equation}
projects it on the fiber of $\psi _{f}$ and then, by unitary evolution, $%
\psi _{f}$ may be reached.\bigskip

(ii) So far, control by measurement plus evolution has been discussed for
finite-dimensional spaces. However the same technique may be used in
infinite-dimensional spaces to reach a large number of states. This is
illustrated for kicked motions in the torus. Let $\overrightarrow{x}=\left(
x_{1},x_{2}\in [-\pi ,\pi )\right) $ be coordinates in the 2-torus $T^{2}$
and the system Hamiltonian $H$ be 
\begin{equation}
H=-\frac{\Delta }{2}+\sum_{n}\left\{ \frac{-1}{2}u_{1}\left( t\right) \left( 
\overrightarrow{x}\bullet A\bullet i\nabla +i\nabla \bullet A\bullet 
\overrightarrow{x}\right) +u_{2}\left( t\right) x_{1}+u_{3}\left( t\right)
x_{2}\right\} \delta \left( t-n\tau \right)  \label{3.15}
\end{equation}
The switching functions $u_{i}\left( t\right) $ take values $0$ or $\pm 1$
and the matrix $A$ is chosen such that $M=\exp \left( A\right) $ is a
hyperbolic 2x2 matrix with integers entries and determinant one, this being
the condition that insures unitarity of the Floquet operator\cite{Vilela2}.
The system is a controlled version of the configurational quantum cat\cite
{Weigert}, a system that describes a charged particle acted upon by
electromagnetic pulses. When $u_{1}\left( t\right) \neq 0$ the Floquet
operator has continuous spectrum and quantum chaos, in the sense of positive
quantum Lyapunov exponents\cite{Vilela1} \cite{Vilela2}. The free and kicked
components of the Floquet operator are 
\begin{equation}
\begin{array}{lll}
U_{0} & = & \exp \left( i\frac{\Delta }{2}\tau \right) \\ 
U_{1} & = & \exp \left( \frac{i}{2}\left( \overrightarrow{x}\bullet A\bullet
i\nabla +i\nabla \bullet A\bullet \overrightarrow{x}\right) \right) \\ 
U_{2} & = & \exp \left( -ix_{1}\right) \\ 
U_{3} & = & \exp \left( -ix_{2}\right)
\end{array}
\label{3.16}
\end{equation}
$U_{0}$ corresponds to free propagation, $U_{1}$ to the action of a linear
vector field and $U_{2},U_{3}$ to scalar potentials. The eigenstates of
momentum form a numerable normalized basis 
\begin{equation}
\left| \overrightarrow{k}\right\rangle =\frac{1}{\sqrt{2\pi }}e^{i%
\overrightarrow{k}\cdot \overrightarrow{x}}\qquad \overrightarrow{k}=\left(
k_{1},k_{2}\right) \qquad k_{i}\in {\Bbb Z}  \label{3.17}
\end{equation}
which is dense on the Hilbert space of the system. The kicks $U_{1}$ act on
these states as follows 
\begin{equation}
U_{1}\left| \overrightarrow{k}\right\rangle =\left| M^{-1}\overrightarrow{k}%
\right\rangle  \label{3.18}
\end{equation}
By the action of kicks of type $U_{1}$ the momentum eigenstates move along
hyperbolas. The kicks $U_{2}$ and $U_{3}$ move between different hyperbola.
In between kicks, free propagation just changes the phase of the states.

Given now an arbitrary state $\psi $ , measuring its momentum the state
becomes projected on a momentum eigenstate with known eigenvalue, if the
result of the measurement is recorded. By switching on the appropriate
sequence of kicks $U_{i}$ it is then possible to reach any momentum
eigenstate. Therefore one sees that with measurement and three controlling
fields one can, from an arbitrary $\psi $, reach any state in an
infinite-dimensional dense set.

\section{Nonlinear and optimal control}

Classical control is a very mature field where many useful techniques and
results have been found, many of them still without parallel in quantum
control. The Strocchi map, yielding a picture of quantum evolution as
Hamiltonian dynamics in a classical-like phase-space, may be the appropriate
tool to carry over techniques from classical to quantum control. We will
give two examples:

(i) {\bf Optimal control}

Optimal control is an important issue both in classical and quantum control
and, in quantum control, it has been discussed using variational techniques%
\cite{optiquan1} \cite{optiquan2} \cite{optiquan3}. However, in classical
control, Pontryagin maximum principle\cite{Pontrya} provides a more general
framework in the sense that it does not require differentiability and can
handle piecewise continuous and magnitude-limited control. We now show how
to carry over this principle to quantum control using the Strocchi map.

In addition to the $2N$ phase-space variables $\left(
x_{k}=q_{k};x_{k+N}=p_{k}\right) $ with dynamical laws 
\begin{equation}
\begin{array}{lllllll}
\frac{d}{dt}x_{k} & = & \frac{d}{dt}q_{k} & = & \frac{\partial }{\partial
p_{k}}{\Bbb H} & = & X_{k} \\ 
\frac{d}{dt}x_{k+N} & = & \frac{d}{dt}p_{k} & = & -\frac{\partial }{\partial
q_{k}}{\Bbb H} & = & X_{k+N}
\end{array}
\label{4.2}
\end{equation}
obtained from (\ref{1.5}), we introduce a variable $x_{0}$ with dynamical
law 
\begin{equation}
\frac{d}{dt}x_{0}=X_{0}\left( x,u,t\right)  \label{4.3}
\end{equation}
with 
\begin{equation}
F=\int_{0}^{T}X_{0}\left( x,u,t\right) dt  \label{4.4}
\end{equation}
being the performance functional to be minimized. If, for example, minimal
controlling energy is desired $X_{0}\left( x,u,t\right) =\left| u\left(
t\right) \right| ^{2}$, etc.

Then, for each variable in the set$\overrightarrow{x}=\left(
x_{0},q_{1},\cdots ,q_{N},p_{1},\cdots ,p_{N}\right) $ an adjoint variable $%
\phi _{i}$ is defined and a new ``Hamiltonian'' ${\bf H}\left( x,\phi
,u\right) $%
\begin{equation}
{\bf H}\left( x,\phi ,u\right) =\sum_{i=0}^{2N}\phi _{i}X_{i}\left(
x,u\right)  \label{4.5}
\end{equation}
For each specified initial $\left(
x_{k}=q_{k}=q_{k}(0),x_{k+N}=p_{k}=p_{k}(0)\right) $and final state $\left(
x_{k}=q_{k}=q_{k}(T),x_{k+N}=p_{k}=p_{k}(T)\right) $, the optimal control
that, in time $T$, minimizes the functional $F$ is obtained by integration
of 
\begin{equation}
\begin{array}{lll}
\frac{d}{dt}x_{k} & = & \frac{\partial }{\partial \phi _{k}}{\bf H}\left(
x,\phi ,u\right) \\ 
\frac{d}{dt}\phi _{k} & = & -\frac{\partial }{\partial x_{k}}{\bf H}\left(
x,\phi ,u\right) \\ 
\overrightarrow{u} & = & {\arg \max }_{\overrightarrow{u}\in \Omega }{\bf H}%
\left( x,\phi ,u\right)
\end{array}
\label{4.6}
\end{equation}
$\Omega $ being the domain of allowed controls.

In the variational formulation of optimal quantum control, the technique
that has been used is to minimize the functional 
\[
J=\left\langle \psi \left( T\right) \right| \left( 1-\Pi \right) \left| \psi
\left( T\right) \right\rangle +\int_{0}^{T}X_{0}\left( \psi ,u,t\right)
dt+\int_{0}^{T}\textnormal{Im}\left\langle \zeta \left( t\right) \right|
i\partial _{t}-H\left| \psi \left( t\right) \right\rangle 
\]
$\Pi $ being the projector onto the target state. The first two equations in
(\ref{4.6}) correspond to the Schr\"{o}dinger equation and to the equation
for the Lagrange multiplier $\zeta \left( t\right) $, whereas the third one
corresponds to the equation obtained by variation $\delta \overrightarrow{u}$
of the control parameters. However the set (\ref{4.6}) is more general in
that it does not require differentiability in $u$ of $J$ and allows the
specification of arbitrary control domains $\Omega $.\bigskip

(ii) {\bf Sliding mode techniques in quantum control}

A very robust tool used in classical control is the technique of variable
structure control leading to sliding modes\cite{Zak}. The design of a
variable structure control has two steps. First, a switching surface must be
chosen so that the dynamical system restricted to the surface has the
desired dynamics. Second, a switched control must be found to drive the
system to the switching surface and, upon interception, to maintain it
there. For this step a Lyapunov function approach is used, the gradient of
the Lyapunov function being negative in the neighborhood of the switching
surface. In this way the tangent vectors to the state trajectory point
towards the surface. The system is attracted to the switching surface and,
once having intercepted it, remains there for all subsequent times. Then,
the state trajectory is said to be a {\it sliding mode}.

From the very nature of the technique one sees that, at least close to the
switching surface, the dynamics must have a dissipative component and
therefore a purely Hamiltonian control cannot be used. However, as seen in
Section 1, observation introduces non-Hamiltonian effects which might play
for quantum dynamics the same role as dissipation in classical control.

Consider again the non-controllable example of the previous section and
suppose that one wants to stabilize the middle energy level. Starting from
an arbitrary initial state one measures the energy. If the result is zero we
are done because it means that the state was projected on the middle level.
Otherwise if the result is $+\mu $ or $-\mu $, one knows from the
controlling subgroups in Eqs.(\ref{3.8a}-\ref{3.10}), that there is a
control that changes the upper or the lower level into a state with a 50\%
probability of being projected on the middle energy state by an energy
measurement. One performs this control plus measurement operation until the
result of the measurement is zero. Afterwards the control is switched off.
If there are no disturbances the state remains in the middle level.
Otherwise if there are some decoherent interaction with the environment, the
state should be periodically measured. If the disturbance is small or if the
intervals between measurements are small, there is a high probability that
the system will be projected back on the middle level, without any need for
further controlling operations.

\section{Conclusions}

In addition to a formulation of quantum control in a symplectic geometry
setting, the main result of this work is the proposal of a new protocol for
quantum control, which we have called ``control by measurement plus
evolution''. It extends the scope of quantum controllability and is
applicable both to finite and infinite-dimensional level systems.

Different aspects of quantum control are united in the framework of the
Strocchi map, allowing new insights in optimal control and nonlinear control
techniques. We think that the potential of this picture is not exhausted and
more analogies with classical problems may be used to obtain progress in the
quantum domain.

\end{document}